\documentclass[10pt,a4paper]{article}
\usepackage[margin=25mm]{geometry}
\usepackage{cite}
\usepackage{amsmath,amssymb,amsthm,mathtools}
\usepackage{microtype}
\usepackage[colorlinks=true,linkcolor=blue,urlcolor=blue,citecolor=blue]{hyperref}
\usepackage{enumitem}

\usepackage{comment}

\usepackage{tikz}
\usetikzlibrary{arrows.meta}
\setlist{nosep}
\numberwithin{equation}{section}
\newtheorem{theorem}{Theorem}
\newtheorem{proposition}[theorem]{Proposition}
\newtheorem{lemma}[theorem]{Lemma}

\newtheorem{corollary}[theorem]{Corollary}

\theoremstyle{definition}
\newtheorem{remark}[theorem]{Remark}
\newtheorem{example}[theorem]{Example}
\newcommand{\R}{\mathbb R}
\newcommand{\ee}{\boldsymbol{e}}
\title{Cycles of Arbitrary Period in Multidimensional\\Heterogeneous Hegselmann–Krause Dynamics}
\author{Anton V. Proskurnikov\thanks{Politecnico di Torino, Turin, Italy, \texttt{anton.p.1982@ieee.org}}\;\; 
and Iryna Zabarianska\thanks{Moscow Institute of Physics and Technology, Moscow, Russia, \texttt{akshiira@yandex.ru}} 
}

\date{}
\begin{document}
\maketitle
\begin{abstract}
We study periodic behavior in heterogeneous Hegselmann--Krause opinion dynamics with
multidimensional opinions and norm-defined confidence neighborhoods. We prove that, for every
dimension $d\ge3$, every $\ell_q$ norm with $q\in[1,\infty]\setminus\{2\}$, and every integer
$m\ge2$, a system of five agents admits a periodic trajectory of minimal period $m$; moreover, different periods can coexist for fixed confidence radii. Four agents remain stationary, while the fifth moves along a line segment contained in their convex hull. The stationary agents may share a common strictly positive confidence radius, and all confidence
inequalities are strict. A geometric criterion underlying the construction extends these results to further classes of norms. However, this criterion cannot be satisfied in dimensions one and two, regardless of the confidence norm. We also prove that no heterogeneous Hegselmann--Krause system with less than five agents admits a nonconstant periodic trajectory, under any norm and in any dimension; every
trajectory with at most three agents converges exponentially. 
\end{abstract}

\section{Introduction}

The Hegselmann--Krause (HK) model~\cite{HegselmannKrause2002,BernardoEtAl2024} describes the 
evolution of opinions in a group of interacting agents and captures the phenomenon of homophily. Individuals in this model adopt opinions
similar to their own while disregarding sufficiently dissimilar ones. Each agent $i$ is characterized by a confidence radius
$\rho_i>0$, which determines the opinions it takes into account. For a group of $n$ agents with opinions $x_i(t)\in\R^d$ and a fixed norm $\|\cdot\|$ on $\R^d$, the HK dynamics is defined by
\begin{equation}\label{eq:hk}
 N_i(t)=\{j:\|x_j(t)-x_i(t)\|\le\rho_i\},\qquad
 x_i(t+1)=\frac{1}{|N_i(t)|}\sum_{j\in N_i(t)}x_j(t)\qquad \forall i=1,\ldots,n.
\end{equation}
Most existing studies focus on scalar opinions and identical confidence radii. However, the possibility of agent-dependent confidence radii was already considered by Hegselmann and Krause~\cite{HegselmannKrause2002}, while
multidimensional extensions of the HK model were proposed a few years later in~\cite{Fortunato2005Vector,Lorenz2006Multidimensional}. To the best of our knowledge, the general model~\eqref{eq:hk},
combining multidimensional opinions with heterogeneous confidence radii and arbitrary norms\footnote{Unlike in the scalar case, there is no canonical choice of norm for measuring similarity between multidimensional opinions.
For instance, Douven and Hegselmann~\cite{DouvenHegselmann2022} consider both Euclidean and $\ell_1$ (Manhattan) distances in a multidimensional bounded-confidence model, illustrating the relevance of non-Euclidean metrics for modeling opinion similarity.}, was first explicitly
proposed in the PhD thesis~\cite{Lorenz2007Thesis}. 

The same thesis also raised the fundamental question of asymptotic behavior: whether opinions necessarily converge
to a limit or may continue changing indefinitely. In particular, Lorenz conjectured~\cite[Conjecture 3.3.2]{Lorenz2007Thesis} convergence under heterogeneous confidence bounds and investigated the emergence of consensus and opinion clusters. For the \emph{homogeneous} HK model, with $\rho_i=\rho>0$, convergence (and, in fact, finite-time termination) is guaranteed even for multidimensional opinions and arbitrary norms~\cite{NedicTouri2012,ProskurnikovTempo2018}.

Simulation-based studies in subsequent years~\cite{Lorenz2010Heterogeneous,MirtabatabaeiBullo2012,FuZhangLi2015}
appeared to support Lorenz's conjecture, although periodic behavior was numerically
observed in heterogeneous HK models with interactions constrained by a lattice~\cite{PerrierSchaweHernandez2024}
and in other modifications of classical HK dynamics~\cite{Chazelle2024,ZabarianskaProskurnikov2024}.
Recently, Hegarty, Ognissanti, and Wedin~\cite{HegartyOgnissantiWedin2026}
disproved several auxiliary conjectures and results concerning heterogeneous HK dynamics, without, however,
finding a non-convergent trajectory.

The convergence conjecture itself remained open for almost two decades, until a recent breakthrough
by Etesami and Touri~\cite{EtesamiTouri2026}, who disproved it by constructing an explicit
periodic orbit of the original scalar ($d=1$) heterogeneous HK model with only six agents.
Although constructed for scalar opinions, their example immediately extends to arbitrary dimensions and norms: indeed, any scalar HK trajectory can be embedded into a line spanned by a unit vector, where the induced norm is the absolute value. 
Their elegant result not only settles a long-standing open problem but also raises new questions about the possible
periods and minimal configurations supporting non-convergent behavior.

In this work, we propose an alternative construction that applies in dimensions $d\geq 3$ and accommodates a broad
class of non-Euclidean norms, including all weighted $\ell_q$ norms with $q\in[1,\infty]\setminus\{2\}$.
Whereas the example of Etesami and Touri~\cite{EtesamiTouri2026} involves a reflection-symmetric configuration of $n=6$ agents, arranged in three pairs with three distinct confidence radii, our construction requires only five agents: four stationary ``anchors'' and a single moving agent whose opinion oscillates along a line segment contained in the convex hull of the anchors. Moreover, all anchors can share the same confidence radius, so that only two distinct radii are required (the minimum possible, since the homogeneous model converges in any norm and any dimension~\cite{ProskurnikovTempo2018}). Furthermore, we show that multidimensional HK dynamics with five agents can exhibit periodic orbits of any prescribed minimal period
$m\geq 2$. This contrasts with the scalar case, where period-2 orbits are impossible, as proved
in~\cite{EtesamiTouri2026}; the counterexample constructed in~\cite{EtesamiTouri2026} has minimal period ten.
As in~\cite{EtesamiTouri2026}, our construction satisfies \emph{strict confidence inequalities}, meaning that
\begin{equation}\label{eq:strict-inequalities}
\|x_j(t)-x_i(t)\|\ne\rho_i    
\end{equation}
for all agents $i,j$ and all times $t$. Consequently, the periodic orbits remain valid when the HK confidence neighborhoods $N_i(t)$ are defined using strict inequalities $\|x_j-x_i\|<\rho_i$. The restriction $d\geq 3$ is essential for the mechanism: in the plane it fails for every norm
(Proposition~\ref{prop:escape-dimension}). Whether five agents admit planar periodic orbits of a different
type remains open, while six agents suffice in every dimension by~\cite{EtesamiTouri2026}.

On the other hand, we prove that, for any norm on a finite-dimensional opinion space, the heterogeneous HK model with $n\leq 4$ admits no nonconstant periodic orbits; moreover, the HK model with $n\leq 3$ agents converges. Thus, five is the minimum number of agents required for periodic behavior in multidimensional heterogeneous HK dynamics, and our construction attains this bound.

Our construction was motivated by the study of multidimensional HK dynamics with stubborn agents~\cite{ChazelleWang2017,ZabarianskaProskurnikov2025}, where open-minded agents with a
common confidence radius interact with stubborn agents (``anchors'') whose opinions remain fixed.
Formally, stubborn agents obey~\eqref{eq:hk} with $\rho_i=0$: their closed confidence
neighborhoods contain only agents with the same opinion. Chazelle and Wang~\cite{ChazelleWang2017}
proved that such dynamics converge when the confidence balls are Euclidean. In contrast, a single
open-minded agent interacting with four anchors can oscillate periodically forever when the
confidence balls are those of a weighted $\ell_q$ norm, $q\ne2$, in dimension $d\ge3$. 
Moreover, the anchors need not be stubborn: sufficiently small positive confidence radii keep them stationary,
and the same periodic trajectories arise in the standard heterogeneous HK model~\eqref{eq:hk}.

The paper is organized as follows. Section~\ref{sec:simple} presents two simple examples,
short cycles in the $\ell_1$ and $\ell_\infty$ norms, that illustrate the main idea.
Section~\ref{sec:general} establishes a geometric escape criterion that guarantees a periodic orbit
of a prescribed minimal period $m\ge2$. In Section~\ref{sec:ell-q} we verify this criterion for all weighted
$\ell_q$ norms with $q\in[1,\infty]\setminus\{2\}$. Section~\ref{sec:n5minimal} proves that $n=5$ is the minimal
number of agents, admitting non-constant periodic cycles; for $n\leq 3$, the HK model converges, while for $n=4$ we prove the absence of nonconstant periodic orbits (technical details of the proof are deferred to a technical Appendix). Section~\ref{sec:discussion} discusses the limitations of our construction and remaining open problems. Section~\ref{sec:conclusion} concludes the paper.

\section{Two Examples of Periodic Trajectories}\label{sec:simple}

Our first example shows that period-two orbits are possible in multidimensional HK dynamics with the $\ell_1$ norm, even though
they are excluded in the scalar heterogeneous model~\cite{EtesamiTouri2026}.

\begin{example}[An exact period-two orbit]\label{ex:two}
Consider five agents in $\R^3$ equipped with the $\ell_1$ norm.
The first four agents have confidence radii $\rho_i=2$
and remain stationary at
\begin{equation}\label{eq:anchors-two}
 x_1=(0,0,0),\quad x_2=(9,0,0),\quad
 x_3=(0,9,0),\quad x_4=(0,0,9).
\end{equation}
The fifth agent has radius $\rho_5=21/2$ and starts at $x_5(0)=\ee:=(1,1,1)$. Its opinion follows the cycle
\begin{equation}\label{eq:twocycle}
 \boxed{\ee\ \longmapsto\ 2\ee\ \longmapsto\ \ee}.
\end{equation}
Indeed, at $x_5=\ee$, the distances to $x_1,x_2,x_3,x_4$
are $3,10,10,10$, respectively, all strictly below $\rho_5$.
Since $x_1+x_2+x_3+x_4=9\ee$, the updated opinion of agent $5$ is
$x_5^+=(\ee+9\ee)/5=2\ee$.
At $x_5=2\ee$, the corresponding distances are
$6,11,11,11$, so only $x_1$ is trusted, and agent $5$ updates its opinion to $x_5^+=(2\ee+x_1)/2=\ee$.
The first four agents always trust only themselves, since their mutual distances and the distances to $\ee,2\ee$ all exceed $2$. Thus, \eqref{eq:twocycle} is an exact orbit of minimal period two. 
\end{example}

Our second example exhibits a similar configuration of four
stationary agents and one moving agent, but with a minimal
period-three orbit under the $\ell_\infty$ norm.

\begin{example}[An exact period-three orbit]\label{ex:three}
Consider five agents in $\R^3$ equipped with the
$\ell_\infty$ norm. The first four agents have confidence
radii $\rho_i=1/2$ and remain stationary at
\begin{equation}\label{eq:anchors-three}
 x_1=(0,0,0),\quad x_2=(-15,17,17),\quad
 x_3=(17,-15,17),\quad x_4=(17,17,-15).
\end{equation}
The fifth agent has confidence radius $\rho_5=33/2$.
Starting at $x_5(0)=\ee=(1,1,1)$, its opinion follows
\begin{equation}\label{eq:threecycle}
 \boxed{\ee\ \longmapsto\ 4\ee\ \longmapsto\
 2\ee\ \longmapsto\ \ee}.
\end{equation}
Indeed, at $x_5=\ee$, the distances to $x_2,x_3,x_4$ are all $16<\rho_5$, and the distance to $x_1$ is $1$.
Since $x_1+x_2+x_3+x_4=19\ee$, averaging gives $x_5^+=(\ee+19\ee)/5=4\ee$.
At $x_5=4\ee$ and $x_5=2\ee$, the distances to each of $x_2,x_3,x_4$ are $19$ and $17$, respectively,
both exceeding $\rho_5$. Meanwhile, $x_1$ remains trusted, so the moving opinion is halved until it reaches $\ee$ again.
The first four agents always trust only themselves, since their mutual distances and their distances to
$\ee,2\ee,4\ee$ are at least $1>1/2$. We have found an exact orbit~\eqref{eq:threecycle} of minimal period three. 
\end{example}

\begin{remark}[Periodic behavior under bounded influence]\label{rem:SBI}
The configuration in Example~\ref{ex:three} also yields a periodic orbit in the multidimensional
``synchronous bounded-influence'' (SBI) model~\cite{MirtabatabaeiBullo2012,HegartyOgnissantiWedin2026SBI}. The opinion update rule is same as in~\eqref{eq:hk}, but the neighborhoods are defined by
\[
N_i(t)=\{j:\|x_j(t)-x_i(t)\|\le\rho_j\},
\]
so that each agent's radius determines which other agents it can influence.

Indeed, retain the initial opinions from Example~\ref{ex:three}, but assign the influence radii
\[
\rho_1=\rho_2=\rho_3=\rho_4=33/2,
\qquad \rho_5=1/2.
\]
The first four agents remain stationary: their mutual distances exceed $33/2$, while
their distances to the moving agent exceed $1/2$ throughout the orbit.
The fifth agent, in turn, is influenced by all four stationary agents at $\ee$, but only
by the agent at the origin at $2\ee$ and $4\ee$. Consequently, the exact period-three
orbit~\eqref{eq:threecycle} persists, with all influence inequalities strict.
Whereas the scalar SBI construction of~\cite{HegartyOgnissantiWedin2026SBI}
involves $53$ agents arranged in five clusters and yields \emph{asymptotically} periodic behavior,
our example exhibits an \emph{exact} period-three
orbit with only five agents in dimension three.
\end{remark}

Both examples rely on the same switching mechanism: the moving agent first averages its opinion with those of all four
stationary agents and then repeatedly averages only with the ``anchor'' at the origin, resulting in successive halvings of its opinion until it returns to the initial state. Remarkably, the moving agent's opinion remains confined to a line segment, although the geometry of the confidence neighborhoods is genuinely multidimensional. In both examples all confidence inequalities are strict, with margin at least $1/2$. This mechanism is not specific to the $\ell_1$ or $\ell_\infty$ norm: we show below that it can be realized with more general norms, yielding periodic orbits of every prescribed minimal period.

In the remainder of the paper, we focus on the bounded-confidence model~\eqref{eq:hk}, which is
the more commonly studied formulation of heterogeneous HK dynamics. Nevertheless, the
geometric construction of the escape mechanism developed in the next section can also be adapted to the multidimensional
bounded-influence model discussed in Remark~\ref{rem:SBI}, provided that additional
separation conditions are imposed to ensure that the anchors remain stationary.

\section{A General Construction}\label{sec:general}

We now generalize the mechanism underlying Examples~\ref{ex:two} and~\ref{ex:three}.
The key is an \emph{escape mechanism}: after averaging with all four stationary
agents, the moving agent loses confidence links to three of them, retaining only
the anchor at the origin. Repeated averaging with this anchor then brings the moving agent
back to its initial opinion.

We initially assume that the first four agents are stubborn, with fixed opinions $x_1,\ldots,x_4\in\R^d$
(equivalently, $\rho_i=0$ in~\eqref{eq:hk}) and denote $\rho_5:=r>0$.

\begin{proposition}[Escape criterion]
\label{prop:escape}
Fix an integer $m\ge2$ and let $H=2^{m-1}$.
Suppose there exist a vector $\ee\in\R^d\setminus\{0\}$, a radius $r>0$ and
vectors $w_1,w_2,w_3\in\R^d$ such that
\begin{gather}
\|w_i\|<r,\qquad \|w_i-\ee\|>r,\quad i=1,2,3,\label{eq:escape-norm}\\
w_1+w_2+w_3=(5H-4)\ee.\label{eq:escape-balance}
\end{gather}

Then, with four stationary agents at
\begin{equation}\label{eq:general-anchors}
x_1=0,\qquad x_{i+1}=\ee+w_i,\quad i=1,2,3,
\end{equation}
the fifth agent with $\rho_5=r$ and $x_5(0)=\ee$ follows the periodic orbit of minimal period $m$:
\begin{equation}\label{eq:general-cycle}
\ee\longmapsto H\ee\longmapsto
\tfrac H2\ee\longmapsto\cdots
\longmapsto 2\ee\longmapsto\ee,
\end{equation}
and this orbit lies in the convex hull of the static opinions $x_1,x_2,x_3,x_4$.
\end{proposition}
\begin{proof}
By the triangle inequality and the balancing condition,
\[
(5H-4)\|\ee\|
\le \sum\nolimits_{i=1}^3\|w_i\|<3r.
\]
Since $5H-4\ge 3H$ for $H\ge2$ we obtain $H\|\ee\|<r$. 

At $x_5=\ee$, all four stationary opinions are trusted: indeed, $\|\ee\|<r$ and
$\|x_{i+1}-\ee\|=\|w_i\|<r$ by~\eqref{eq:escape-norm}. Hence, starting at $\ee$, the moving agent updates its opinion to
\[
x_5^+
=\frac{4\ee+\sum_{i=1}^3w_i}{5}
=H\ee.
\]

Consider now $x_5=h\ee$, where $2\le h\le H$. The agent at the origin remains trusted, since $\|h\ee\|\leq H\|\ee\|<r$.
The other three stationary agents are not trusted by agent $5$.
Indeed, if $\|x_{i+1}-h\ee\|=\|w_i-(h-1)\ee\|\leq r$, then the convexity of the norm would imply
\[
\|w_i-\ee\|
=\left\|\frac{h-2}{h-1}w_i
+\frac{1}{h-1}\bigl(w_i-(h-1)\ee\bigr)\right\|\leq r
\]
contrary to~\eqref{eq:escape-norm}. Hence $\|x_{i+1}-h\ee\|>r$, and at every such state the fifth agent
averages only with the origin and itself, giving
$x_5^+=h\ee/2$. Hence, the opinion of agent $5$ evolves as in~\eqref{eq:general-cycle};
its $m$ possible states are distinct since $\ee\ne0$, so $m$ is the minimal period.
The statement about the convex hull follows from the fact that $5H-1>3H$, and thus every $h\ee$ with $1\leq h\leq H$ belongs to the line segment connecting $x_1=0$ to the point $(x_2+x_3+x_4)/3=(5H-1)/3\,\ee$.
\end{proof}

\begin{corollary}\label{cor:positive-radii}
Under the assumptions of Proposition~\ref{prop:escape}, let $\delta>0$ be the smallest distance between
distinct positions occupied by the five agents along the orbit~\eqref{eq:general-cycle}.
If the first four agents are ordinary HK agents with radii $\rho_1,\ldots,\rho_4\in(0,\delta)$
(for instance, a common radius $\rho\in(0,\delta)$), then they remain stationary
and~\eqref{eq:general-cycle} is an orbit of~\eqref{eq:hk}. Moreover, all confidence inequalities are strict, i.e.,~\eqref{eq:strict-inequalities} holds, so the same orbit exists when confidence balls are open.
\end{corollary}
\begin{proof}
Each of the first four agents is at distance $0<\rho_i$ from the agents sharing its position and at distance
at least $\delta>\rho_i$ from all other agents. Hence it averages only over agents at its own position and
does not move, and all confidence inequalities involving it are strict. For agent~5, strictness follows
from the proof of Proposition~\ref{prop:escape}: throughout the orbit, the distance to the origin is
strictly less than $r$, while the other three distances are less than $r$ when $x_5=\ee$ and
strictly greater than $r$ at all other states.
\end{proof}

Examples~\ref{ex:two} and~\ref{ex:three} are particular cases of the escape criterion. In both cases,
$\ee=(1,1,1)$ and $w_i=x_{i+1}-\ee$, $i=1,2,3$. For Example~\ref{ex:two}, we have $H=2$ and
\[
\|w_i\|_1=10<\tfrac{21}{2}<11=\|w_i-\ee\|_1,
\qquad \sum\nolimits_{i=1}^3 w_i=6\ee.
\]
For Example~\ref{ex:three}, we have $H=4$ and
\[
\|w_i\|_\infty=16<\tfrac{33}{2}<17=\|w_i-\ee\|_\infty,
\qquad \sum\nolimits_{i=1}^3 w_i=16\ee.
\]
Thus, both examples arise from the same geometric
mechanism, despite using different norms.

We will use the escape criterion to construct periodic orbits of arbitrary minimal period for a broad class of
non-Euclidean norms. Notably, the criterion \emph{cannot} be satisfied for norms induced by an inner product. As will be discussed, this is consistent with the results of~\cite{ChazelleWang2017}.
\begin{remark}\label{rem:no-Euclidean}
The escape criterion cannot be satisfied with any $m\geq 2$ and any norm induced by an inner product $\langle\cdot,\cdot\rangle$. Indeed, it is required that $\|w_i-\ee\|^2>r^2>\|w_i\|^2$, and hence
$2\langle w_i,\ee\rangle
<\|\ee\|^2$ for all $i=1,2,3$. Summing these inequalities and using
$\sum_{i=1}^3w_i=(5H-4)\ee$, we obtain $2(5H-4)\|\ee\|^2
<3\|\ee\|^2$, contrary to $H\ge2$ and $\ee\ne0$.
\end{remark}

Finally, our construction extends to the set-based HK model of~\cite{ZabarianskaProskurnikov2024}, where the agents use general confidence sets, that is, $N_i(t)=\{j:x_j(t)-x_i(t)\in\mathcal{O}_i\}$, where $\mathcal{O}_i\subset\R^d$ are some sets.

\begin{remark}
The escape criterion extends to the set-based HK model in which the neighbors of the fifth agent are
determined by a closed convex set $\mathcal{O}_5=\mathcal{O}$, with~\eqref{eq:escape-norm} replaced by the conditions
\begin{equation}\label{eq:escape1}
-H\ee\in\operatorname{int}\mathcal O,\qquad 0\in \operatorname{int}\mathcal O,\qquad
w_i\in\operatorname{int}\mathcal O,\qquad
w_i-\ee\notin\mathcal O,\quad i=1,2,3
\end{equation}
Indeed, by convexity,~\eqref{eq:escape1} implies that $-h\ee\in\operatorname{int}\mathcal O$
for $0\le h\le H$, so the agent at the origin remains trusted. At $x_5=\ee$, all stationary opinions are
trusted, since $x_{i+1}-x_5=w_i\in\operatorname{int}\mathcal{O}$ for $i=1,2,3$. On the other hand, for
$2\le h\le H$ none of the stationary opinions $x_2,x_3,x_4$ is trusted at $x_5=h\ee$, since
$w_i-(h-1)\ee\notin\mathcal{O}$. Indeed, if $w_i-(h-1)\ee\in\mathcal{O}$, then, since $w_i\in\mathcal{O}$,
\[
w_i-\ee
=\frac{h-2}{h-1}\,w_i
+\frac{1}{h-1}\bigl(w_i-(h-1)\ee\bigr)
\]
would also belong to $\mathcal{O}$ by convexity. Hence the proof of Proposition~\ref{prop:escape} carries over.
 
Corollary~\ref{cor:positive-radii} also remains valid. Fix any norm on $\R^d$ and let $\delta>0$ be
defined as in Corollary~\ref{cor:positive-radii} with respect to this norm. Then
$\mathcal O_1,\ldots,\mathcal O_4$ can be any neighborhoods of $0$ contained in a ball centered at $0$
of radius smaller than $\delta$. Moreover, since $\mathcal O$ is closed and~\eqref{eq:escape1} holds, no relative position
$x_j-x_i$ along the orbit lies on the boundary of the corresponding confidence set $\mathcal{O}_i$. Hence the same
orbit exists for any confidence sets $\mathcal O_i'$ with
$\operatorname{int}\mathcal O_i\subseteq\mathcal O_i'\subseteq\overline{\mathcal O_i}$,
in particular for the interiors $\operatorname{int}\mathcal O_i$.
\end{remark}

\section{Every period and every non-Euclidean $\ell_q$ norm}\label{sec:ell-q}

We now formulate our main theorem.
\begin{theorem}[Cycles of arbitrary period]
\label{thm:ellq}
Let $d\ge3$, $q\in[1,\infty]\setminus\{2\}$, and $m\ge2$ be an integer. Then there exist five strictly positive confidence
radii $\rho_1,\ldots,\rho_5$ and initial opinions $x_1(0),\ldots,x_5(0)\in\R^d$ such that
the heterogeneous HK model~\eqref{eq:hk} with the $\ell_q$ norm admits a periodic
trajectory of minimal period $m$.

Along this trajectory, the first four agents remain stationary and may have the same confidence radius $\rho_1=\ldots=\rho_4$,
while the fifth agent moves along a line segment contained in the convex hull of their opinions.
All confidence inequalities are strict.
\end{theorem}
\begin{proof}
It suffices to consider $d=3$, since the
construction extends to higher dimensions by
appending zero coordinates. Fix the desired minimal period $m\ge2$ and put
\[
H=2^{m-1},\qquad c=\frac{5H-4}{3}.
\]
In order to apply the escape criterion (Proposition~\ref{prop:escape}), let
\[
w_i=Ku_i+c\ee,\qquad i=1,2,3,
\]
where $K>0$ will be chosen sufficiently large, while $\ee$ and $u_i$ are defined as follows:
\begin{gather*}
\ee:=\sigma(1,1,1),\quad \sigma:=
\begin{cases}
-1,&1\le q<2,\\
 1,&2<q\le\infty,
\end{cases}\\
u_1:=(-2,1,1),\quad
u_2:=(1,-2,1),\quad
u_3:=(1,1,-2).
\end{gather*}

The balancing condition~\eqref{eq:escape-balance} is satisfied, since $u_1+u_2+u_3=0$, and thus
\[
\sum\nolimits_{i=1}^3w_i=3c\ee=(5H-4)\ee.
\]

We claim that, for sufficiently large $K$,
\begin{equation}\label{eq:ellq-separation}
\|w_i\|_q<\|w_i-\ee\|_q,
\qquad i=1,2,3.
\end{equation}

Suppose first that $1\leq q<\infty$.
By permutation symmetry, all vectors $w_i$ have the same norm, and
\[
\begin{aligned}
\|w_i\|_q^q
  &=(2K-\sigma c)^q
    +2(K+\sigma c)^q,\\
\|w_i-\ee\|_q^q
  &=(2K-\sigma(c-1))^q
    +2(K+\sigma(c-1))^q,
\end{aligned}
\]
provided $K$ is sufficiently large. 
To compare these expressions, let $F_K(t)=(2K-\sigma t)^q+2(K+\sigma t)^q$. For fixed $t$, the
expansion $(1+s)^q=1+qs+O(s^2)$ as $s\to0$ gives, as $K\to\infty$,
\begin{equation}\label{eq:FK-expansion}
F_K(t)=(2^q+2)K^q+\sigma q\,(2-2^{q-1})\,t\,K^{q-1}+O(K^{q-2}).
\end{equation}
By noticing that $\sigma=\operatorname{sgn}(q-2)=\operatorname{sgn}(2^{q-1}-2)$, one shows that
\[
\|w_i-\ee\|_q^q-\|w_i\|_q^q=F_K(c-1)-F_K(c)=\sigma q\,(2^{q-1}-2)K^{q-1}+O(K^{q-2})
=q\,|2^{q-1}-2|\,K^{q-1}+O(K^{q-2}),
\]
Since $q\ne2$, the leading coefficient is positive, so~\eqref{eq:ellq-separation} holds for all sufficiently large $K$.

For $q=\infty$, we have $\sigma=1$ and, when $K$ is sufficiently large,
\[
\|w_i\|_\infty=2K-c<\|w_i-\ee\|_\infty=2K-c+1,
\]
so \eqref{eq:ellq-separation} holds as well.

Consequently, by permutation symmetry, we can choose a common $r>0$ such that
\[
\|w_i\|_q<r<\|w_i-\ee\|_q,
\qquad i=1,2,3.
\]
All conditions of Proposition~\ref{prop:escape} are thus satisfied.

Thus, with four stationary agents~\eqref{eq:general-anchors}, the fifth agent admits the periodic orbit~\eqref{eq:general-cycle}
whose minimal period is $m$, and this orbit lies in the convex hull of the stationary agents. The assertion concerning strictly positive confidence radii of the static agents and strict inequalities follows from Corollary~\ref{cor:positive-radii}.
\end{proof}

\begin{corollary}
\label{cor:transformed-ellq}
The conclusion of Theorem~\ref{thm:ellq}
remains valid for every norm of the form
\[
\|x\|=\|Tx\|_q,
\]
where $T\in\R^{d\times d}$ is invertible and $q\in[1,\infty]\setminus\{2\}$. In particular, this covers the weighted norms $(\sum_k\omega_k|x_k|^q)^{1/q}$ and $\max_k\omega_k|x_k|$ with $\omega_k>0$ (take $T$ diagonal).
\end{corollary}
\begin{proof}
Apply Theorem~\ref{thm:ellq} in the
coordinates $y=Tx$. Since $T$ is
invertible and linear, averaging and
confidence neighborhoods are preserved
under this transformation.
\end{proof}

\begin{corollary}[Coexistence of different periods]\label{cor:coexistence}
Let $d\ge3$ and $q\in[1,\infty]\setminus\{2\}$. For every integer $M\ge2$, there exist
strictly positive confidence radii $\rho_1=\cdots=\rho_4$ and $\rho_5$ such that the same heterogeneous HK model with the $\ell_q$ norm admits periodic trajectories of every minimal period $m=2,\ldots,M$, for different initial configurations. The same conclusion holds for the norms $x\mapsto\|Tx\|_q$, where $T\in\R^{d\times d}$ is invertible.
\end{corollary}

\begin{proof}
For each $m=2,\ldots,M$, Theorem~\ref{thm:ellq} provides a periodic configuration with
four stationary agents and a fifth agent having confidence radius $r_m>0$.

By homogeneity of the norm, scaling all opinions in this configuration by $1/r_m$
preserves the periodic orbit if the confidence radius of the fifth agent is set to $1$.
Thus, we may take $\rho_5=1$ for every $m$. By Corollary~\ref{cor:positive-radii}, for each of these finitely many configurations
there exists $\varepsilon_m>0$ such that the four stationary agents can all be assigned
the common confidence radius $\varepsilon_m$, without changing the orbit.
Choosing
\[
0<\varepsilon<
\min_{2\le m\le M}\varepsilon_m,
\]
we conclude that the fixed confidence radii
\[
\rho_1=\cdots=\rho_4=\varepsilon,
\qquad \rho_5=1
\]
admit periodic trajectories of every
minimal period $m=2,\ldots,M$,
depending on the initial configuration.

The extension to norms of the form $\|Tx\|_q$ follows by the invertible change of coordinates
used in the proof of Corollary~\ref{cor:transformed-ellq},
which preserves the HK dynamics and the confidence radii.
\end{proof}

\section{Absence of Periodic Orbits for $n\leq 4$ Agents}\label{sec:n5minimal}

The existence of periodic trajectories with five agents contrasts with the absence of nonconstant periodic
orbits for populations of at most four agents. For at most three agents, we obtain the stronger
result of exponential convergence. 

\subsection{Convergence in the case of $n\leq 3$ agents}

We start with the observation that the HK model with $n=2$ agents reaches an equilibrium in finitely many steps (while $n=1$ is trivial).
\begin{lemma}[Finite-time convergence for two agents]
\label{lem:n2}
For the HK model~\eqref{eq:hk} with $n=2$, arbitrary confidence radii
$\rho_1,\rho_2>0$, and any norm on $\R^d$,
every trajectory reaches an equilibrium in
finite time.
\end{lemma}
\begin{proof}
Let $d(t)=\|x_1(t)-x_2(t)\|$.
Depending on whether the agents trust each other,
the distance evolves as follows:
\[
d(t+1)=
\begin{cases}
d(t),&\text{if neither agent trusts the other},\\
d(t)/2,&\text{if exactly one agent trusts the other},\\
0,&\text{if both agents trust each other}.
\end{cases}
\]
If neither agent trusts the other, the trajectory is already stationary. Otherwise, $d(t)$ is at least halved at every step until it becomes smaller than $\min\{\rho_1,\rho_2\}>0$. At the next step, both agents trust each other
and reach consensus. Hence the trajectory reaches an equilibrium in finite time.
\end{proof}

\begin{theorem}[Convergence for three agents]\label{thm:n3-exponential}
Consider heterogeneous HK dynamics with $n=3$, arbitrary strictly positive confidence radii,
and an arbitrary norm on a finite-dimensional opinion space. Then every trajectory converges
exponentially. More precisely, every trajectory either reaches an equilibrium in finite time or, from some time
onward, two agents remain stationary at opinions $a,b$, while the third satisfies
\[
x_i(t+1)=\frac{x_i(t)+a+b}{3}.
\]
In the latter case, $x_i(t)$ converges to $(a+b)/2$ with geometric ratio $1/3$.
\end{theorem}

The proof of Theorem~\ref{thm:n3-exponential} relies on the following technical lemma.

\begin{lemma}[Persistence of full confidence]\label{lem:n3-full}
Consider heterogeneous HK dynamics with $n=3$, arbitrary positive confidence radii, and any norm.
If $N_i(t_0)=\{1,2,3\}$, then $N_i(t)=\{1,2,3\}$ for all $t\geq t_0$.
\end{lemma}
\begin{proof}
We proceed by induction on $t$. The induction base $t=t_0$ is given. For the induction step,
suppose that $N_i(t)=\{1,2,3\}$ and, without loss of generality, let $i=1$.
By translating the opinions, we may also assume that $x_1(t)=0$. Denoting $u:=x_2(t)$ and $v:=x_3(t)$, one thus has
$\|u\|,\|v\|\le\rho_1$. Since agent $1$ trusts both other agents, $x_1(t+1)=(u+v)/3$.
Hence
\[
\begin{aligned}
\|x_1(t+1)-x_1(t)\|
&\le \frac23\rho_1,\\
\|x_1(t+1)-x_2(t)\|
&\le\frac{2\|u\|+\|v\|}{3}
\le\rho_1,\\
\|x_1(t+1)-x_3(t)\|
&\le\rho_1.
\end{aligned}
\]
Thus, the ball centered at $x_1(t+1)$ with radius $\rho_1$ contains all three opinions
at time $t$. By convexity, it also contains their convex hull, and hence all three opinions
$x_i(t+1)$ for $i=1,2,3$. 
\end{proof}

\begin{proof}[Proof of Theorem~\ref{thm:n3-exponential}]
We distinguish two cases.

\emph{Case 1: For all $i=1,2,3$ and $t\geq 0$ one has $N_i(t)\subsetneq\{1,2,3\}$, that is, no agent ever trusts everyone.} 

We will prove that the dynamics terminates in finite time. This case splits into two subcases.

\emph{Case 1a: Two agents trust each other at some time $t$.}
Without loss of generality, let agents $1$ and $2$ trust each other. By assumption,
neither trusts agent $3$, and hence
\[
x_1(t+1)=x_2(t+1)
=\frac{x_1(t)+x_2(t)}2.
\]
At time $t+1$, agents $1$ and $2$ trust each other but cannot trust agent $3$. Since their opinions coincide, agent $3$ cannot trust either of them without trusting both, which would contradict the assumption. Thus the trajectory reaches an equilibrium at time $t+1$.

\emph{Case 1b. No two distinct agents ever trust each other mutually.}
We prove that this is possible only if $N_i(t)\equiv\{i\}$ for every agent $i$, which means that the HK model is at an equilibrium.

Let $V\subseteq\{1,2,3\}$ be the set of agents $j$ such that $N_j(t)\ne\{j\}$ for some $t$.
Suppose that $V\ne\emptyset$, and choose an agent $i\in V$ with the smallest confidence radius.

Suppose that $i$ trusts another agent $j$
at time $t$. Then $j\notin V$: otherwise,
$\rho_j\ge\rho_i$, implying that $j$ also
trusts $i$, a contradiction.
Consequently, agent $j$ remains stationary,
$x_j(t)\equiv x_j$.

Since full confidence is excluded,
$N_i(t)=\{i,j\}$, and hence
\[
x_i(t+1)-x_j=\frac12(x_i(t)-x_j).
\]
Thus, agent $i$ continues to trust $j$.
Since it cannot trust the third agent,
the same argument applies recursively, yielding
$\|x_i(t+k)-x_j\|
=2^{-k}\|x_i(t)-x_j\|$ for $k\ge0$.
Since $\rho_j>0$, agent $j$ eventually trusts
$i$, contradicting the absence of mutual trust.

Therefore, $V=\emptyset$, meaning that
$N_i(t)=\{i\}$ for all $i$ and $t$.

\vskip2mm

\emph{Case 2: Some agent trusts everyone at some time.}
By Lemma~\ref{lem:n3-full}, we may assume, without loss
of generality, that $N_1(t)=\{1,2,3\}$ for all $t\ge T$.
Consider the convex hull of the opinions and its diameter:
\[
C(t)=\operatorname{conv}\{x_1(t),x_2(t),x_3(t)\},
\qquad
D(t)=\max_{i,j}\|x_i(t)-x_j(t)\|.
\]
Since $C(t+1)\subseteq C(t)$, the sequence $D(t)$ is nonincreasing and hence has
a limit $D_*\ge0$.

If $D_*=0$, then, since all confidence radii
are strictly positive, every agent eventually
trusts everyone, and consensus is reached
in finite time.

Suppose henceforth that $D_*>0$. We will prove that agents $2$ and $3$
eventually trust only themselves. The key observation is that whenever, at some
$t\ge T+1$, agent $2$ or $3$ trusts another
agent, one has
\begin{equation}\label{eq:n3-contraction}
D(t+1)\le\frac89D(t-1).
\end{equation}

Indeed, suppose first that agent $i\in\{2,3\}$
trusts agent $1$, and let $k$ denote the remaining
agent. Since agent $1$ trusts everyone,
\[
x_1(t)=\frac{x_1(t-1)+x_2(t-1)+x_3(t-1)}3.
\]
Both agents $1$ and $i$ assign weight at least
$1/3$ to $x_1(t)$ in their next updates.
Hence their opinions at time $t+1$ are convex
combinations of the opinions at time $t-1$,
assigning weight at least $1/9$ to $x_k(t-1)$.
The same holds for agent $k$, since it assigns
weight at least $1/3$ to its own opinion at
each of the two consecutive steps. Since $x_k(t-1)\in C(t-1)$, any convex
combination assigning it weight at least $1/9$
can also be written with weight exactly $1/9$,
absorbing the excess weight into another point
of $C(t-1)$. Thus, 
\[
x_j(t+1)=\frac19x_k(t-1)+\frac89y_j,
\qquad y_j\in C(t-1),\quad\forall j=1,2,3.
\]
The inequality~\eqref{eq:n3-contraction} now follows from the inclusion
\[
C(t+1)\subseteq
\frac19x_k(t-1)+\frac89 C(t-1).
\]

Suppose now that neither agent $2$ nor
agent $3$ trusts agent $1$ at time $t$.
Then one of them, say agent $i$, must trust
the other, denoted by $k$.
Since agent $1$ trusts everyone,
agent $i$ trusts agent $k$, and agent $k$ trusts
itself, all three opinions at time $t+1$ are
convex combinations of the opinions at time $t$
with a common weight at least $1/3$ assigned
to $x_k(t)$. Consequently,
\[
C(t+1)\subseteq
\frac13x_k(t)+\frac23 C(t)\Longrightarrow
D(t+1)\le\frac23D(t)
\le\frac23D(t-1),
\]
proving~\eqref{eq:n3-contraction}, since $2/3<8/9$.

Suppose that agent 2 or agent 3 trusts another agent at infinitely many times.
Passing to the limit along such a sequence in~\eqref{eq:n3-contraction}, we obtain
\[
D_*\le\frac89D_*,
\]
contradicting $D_*>0$.

Therefore, there exists $T_1\ge T$ such that
$N_2(t)=\{2\}$ and $N_3(t)=\{3\}$
for all $t\ge T_1$. Their opinions remain
stationary, $x_2(t)\equiv x_2$ and
$x_3(t)\equiv x_3$, while agent $1$
continues to trust both. Hence,
\[
x_1(t+1)=\frac{x_1(t)+x_2+x_3}{3},
\]
which converges exponentially fast to $(x_2+x_3)/2$. This finishes the proof.
\end{proof}

\begin{remark}[Open confidence balls]\label{rem:n3-open}
Lemmas~\ref{lem:n2}, \ref{lem:n3-full} and Theorem~\ref{thm:n3-exponential} remain valid when
the neighborhoods are defined by $\|x_j-x_i\|<\rho_i$; each agent still trusts itself since
$\rho_i>0$. The radii enter the proofs only in three places, and each argument survives with $<$
in place of $\le$: in Lemma~\ref{lem:n2} the trusting agent keeps trusting since the distance
decreases, and both trust each other once $d(t)<\min\{\rho_1,\rho_2\}$; in
Lemma~\ref{lem:n3-full}, $\|u\|,\|v\|<\rho_1$ yields $\|x_1(t+1)-x_k(t)\|<\rho_1$ for
$k=1,2,3$, and open balls are convex; in Case~1b,
$\|x_i(t)-x_j(t)\|<\rho_i\le\rho_j$ forces $j$ to trust $i$. The remaining steps remain unchanged.

For open balls, moreover, the limit is always an equilibrium. In the second alternative, let
the moving agent have radius $\rho$ and the stationary agents at $a\ne b$ have radii
$\rho_a,\rho_b$, and put $c=(a+b)/2$. The inequalities $\|x_i(t)-a\|\ge\rho_a$ and
$\|x_i(t)-b\|\ge\rho_b$ pass to the limit, so neither stationary agent trusts $c$. Also
$\|c-a\|=\|c-b\|=\|a-b\|/2\le\rho$, and equality is impossible, since then
$\|x_i(t)-a\|+\|x_i(t)-b\|<2\rho=\|a-b\|$. Hence $(c,a,b)$ is an equilibrium. For closed balls
this may fail: with $d=1$, $x(0)=(0.3,-1,1)$ and $\rho=(2,1,\tfrac12)$, one has
$x_1(t)=0.3\cdot3^{-t}\to0$, while agent~2 never trusts agent~1. Yet in the limit
configuration $(0,-1,1)$ it would, since $|0-(-1)|=\rho_2$.
\end{remark}

\subsection{$n=4$ case: no periodic orbits}

While our analysis does not establish convergence of all trajectories for $n=4$, it does exclude
nonconstant periodic solutions, under any norm and in any dimension.

\begin{theorem}[Nonexistence of periodic orbits for $n=4$]
\label{thm:n4}
For heterogeneous HK dynamics with $n=4$ agents,
arbitrary positive confidence radii, and any norm
on $\R^d$, every periodic trajectory is stationary.
\end{theorem}

The proof of the theorem is given in the Appendix.

\section{Discussion and Open Problems}\label{sec:discussion}

Our results raise several questions about the role of
the opinion space dimension, the geometry of confidence
neighborhoods, and the minimal configurations supporting
periodic behavior. We first discuss the limitations of
our escape mechanism and its relation to HK models with
stubborn agents, and then outline some open problems.

\subsection{Dimension three is minimal for the escape criterion to hold}

Whether heterogeneous HK dynamics with $n=5$ agents admits nonconstant periodic trajectories in
dimensions $d=1$ or $d=2$ is not addressed in this paper. We show, however, that the escape
criterion underlying our construction cannot be satisfied in these dimensions, regardless of the
confidence norm. It suffices to consider $d=2$, since any one-dimensional configuration can be embedded into a normed plane.

\begin{proposition}[Dimension obstruction]
\label{prop:escape-dimension}
Let $\|\cdot\|$ be an arbitrary norm on $\R^2$.
Suppose that vectors $\ee,w_1,w_2,w_3\in\R^2$
and scalars $\xi,r>0$ satisfy the relations
\[
w_1+w_2+w_3=\xi\ee,\qquad
\|w_i\|<r<\|w_i-\ee\|,\quad i=1,2,3.
\]
Then necessarily $\xi<3$. In particular, the escape conditions \eqref{eq:escape-norm} and
\eqref{eq:escape-balance} cannot hold in dimension two with $H=2^{m-1}\ge2$, because
$\xi=5H-4\ge6$.
\end{proposition}
\begin{proof}
Changing the coordinates, one may assume that $\ee=(1,0)$. Writing $w_i=(a_i,b_i)$, one has
\begin{equation}\label{eq:balance-aux}
\sum\nolimits_{i=1}^3a_i=\xi,\qquad
\sum\nolimits_{i=1}^3b_i=0.
\end{equation}
After relabeling and, if necessary, reversing the vertical axis, we may assume $b_1,b_2\ge0$. 

Let $B=\bar B(0,r)$ be the closed confidence ball, and let $I\subset\R$ denote its projection
onto the vertical axis. For every $y\in I$, the horizontal section of $B$ is a (possibly
a one-point) line segment with endpoints $(L(y),y)$ and $(R(y),y)$, where
\[
L(y):=\min\{x:(x,y)\in B\}\leq R(y):=\max\{x:(x,y)\in B\}\quad\forall y\in I.
\]
By convexity and central symmetry of $B$, the function $L:I\to\R$ is convex, and
\begin{equation}\label{eq:L-R-relation}
L(-y)=-R(y)\le-L(y), \qquad y\in I,
\end{equation}
also, one has $L(0)<0$, because $0\in\operatorname{int} B$. 

Since $w_i=(a_i,b_i)\in B$, one has $b_i\in I$ and $a_i\in[L(b_i),R(b_i)]$. On the other hand, $w_i-\ee=(a_i-1,b_i)\notin B$, and thus
$a_i-1<L(b_i)$. Denote $s:=b_1+b_2\geq 0$. Since $(-s)=b_3\in I$ and $I$ is symmetric, one has $s\in I$. If $s>0$, convexity gives
\[
L(b_1)\le\frac{b_2}{s}L(0)
          +\frac{b_1}{s}L(s),\qquad
L(b_2)\le\frac{b_1}{s}L(0)
          +\frac{b_2}{s}L(s).
\]
Adding these inequalities yields $L(b_1)+L(b_2)\leq L(0)+L(s)$, which inequality also holds when $s=0$ (i.e., $b_1=b_2=0$). 
The relations~\eqref{eq:balance-aux} and~\eqref{eq:L-R-relation} imply that
\[
\begin{aligned}
\xi=\sum\nolimits_{i=1}^3a_i
 <3+\sum\nolimits_{i=1}^3L(b_i)\le3+L(0)+L(s)+L(-s)\le3+L(0)<3,
\end{aligned}
\]
which finishes the proof of Proposition~\ref{prop:escape-dimension}.
\end{proof}

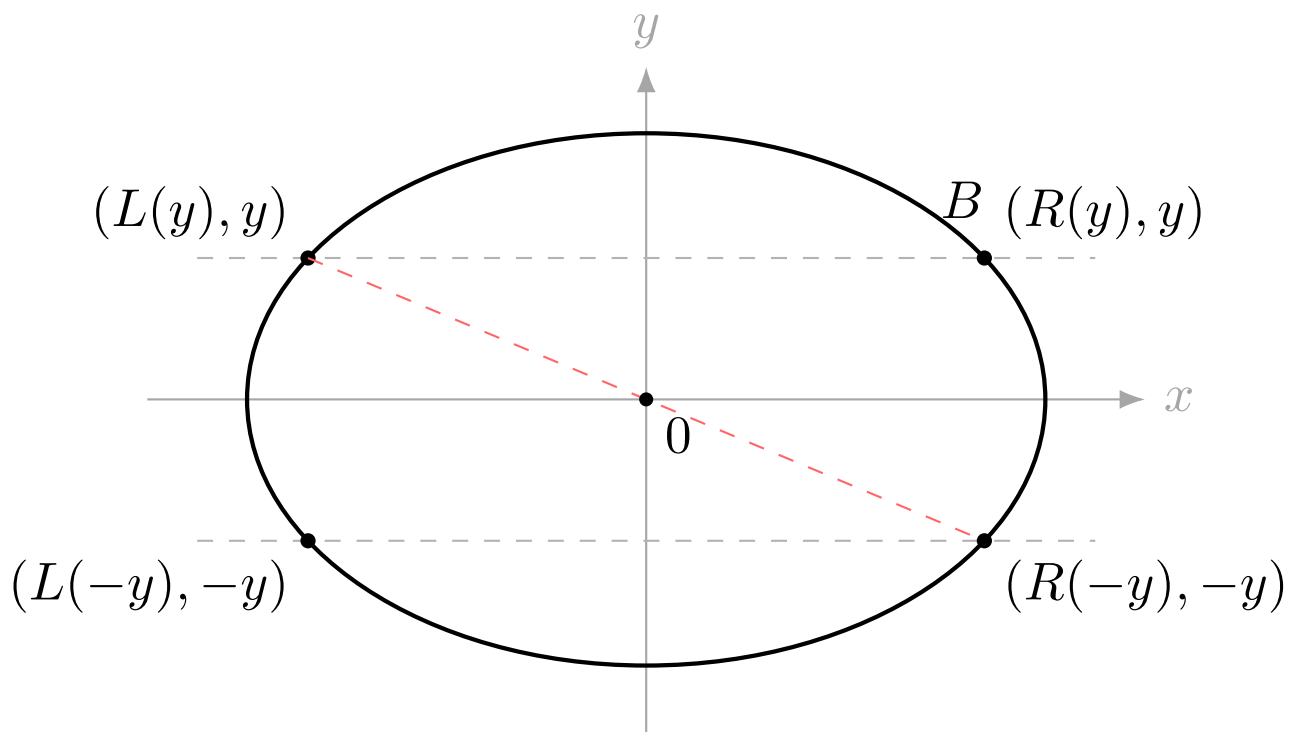
\begin{figure}[t]
\centering
\begin{tikzpicture}[scale=1.1,>=Latex]

\def\a{2.4}
\def\b{1.6}
\def\yy{0.85}

\pgfmathsetmacro{\Ly}{-\a*sqrt(1-(\yy/\b)^2)}
\pgfmathsetmacro{\Ry}{ \a*sqrt(1-(\yy/\b)^2)}
\pgfmathsetmacro{\Lmy}{-\a*sqrt(1-((-\yy)/\b)^2)}
\pgfmathsetmacro{\Rmy}{ \a*sqrt(1-((-\yy)/\b)^2)}

\draw[->,gray!70] (-3.0,0) -- (3.0,0) node[right] {$x$};
\draw[->,gray!70] (0,-2.0) -- (0,2.0) node[above] {$y$};

\draw[thick] (0,0) ellipse ({\a} and {\b});
\node at (1.9,1.2) {$B$};

\draw[dashed,gray!60] (-2.7,\yy) -- (2.7,\yy);
\draw[dashed,gray!60] (-2.7,-\yy) -- (2.7,-\yy);

\fill (\Ly,\yy) circle (1.3pt);
\fill (\Ry,\yy) circle (1.3pt);
\fill (\Lmy,-\yy) circle (1.3pt);
\fill (\Rmy,-\yy) circle (1.3pt);

\draw[dashed,red!60](\Ly,\yy) -- (\Rmy,-\yy);

\node[above left] at (\Ly,\yy) {$(L(y),y)$};
\node[above right] at (\Ry,\yy) {$(R(y),y)$};
\node[below left] at (\Lmy,-\yy) {$(L(-y),-y)$};
\node[below right] at (\Rmy,-\yy) {$(R(-y),-y)$};

\fill (0,0) circle (1.2pt);
\node[below right] at (0,0) {$0$};
\end{tikzpicture}

\caption{Schematic illustration of the left and right boundary functions $L,R$ of a centrally symmetric convex body $B\subset\mathbb R^2$. For each height $y$, the horizontal section of $B$ is the interval $[L(y),R(y)]$. Central symmetry implies $L(-y)=-R(y)$ (the red dashed line) for each $y$.}
\label{fig:left-right-boundaries}
\end{figure}

\subsection{Connection with HK models with stubborn agents}

Our construction is closely related to HK models with stubborn agents, whose opinions remain fixed
regardless of their neighbors. Indeed, the first four agents in our periodic trajectories are
stationary and can equivalently be treated as stubborn agents with zero confidence radii. 

Theorem~\ref{thm:ellq} provides a negative answer to the open problem formulated in~\cite{ZabarianskaProskurnikov2025}:
for every $\ell_q$ norm with $q\in[1,\infty]\setminus\{2\}$ and $d\ge3$, the HK model with four stubborn agents and
a single open-minded agent can fail to converge and even exhibit nonconstant periodic trajectories.
Moreover, our construction shows that these oscillations persist when the stubborn agents are replaced by regular HK agents with sufficiently small, strictly positive confidence radii. Thus, convergence under a general norm cannot be guaranteed without
additional assumptions on the configuration of stubborn opinions. Sufficient conditions established
in~\cite{ZabarianskaProskurnikov2025} include the case where all stubborn opinions are sufficiently close (e.g., only one stubborn agent is present~\cite{ZabarianskaProskurnikov2024}), as well as the case
where they form tight clusters separated by sufficiently large distances.

This observation is particularly interesting in view of the convergence results established by Chazelle
and Wang~\cite{ChazelleWang2017} for Euclidean HK models with stubborn agents, in which all open-minded
agents have a common confidence radius $\rho_*>0$. The impossibility of our escape mechanism for
Euclidean confidence balls, established in Remark~\ref{rem:no-Euclidean}, is consistent with these convergence results.
A key ingredient of their analysis is the finiteness
of the kinetic $2$-energy for inertial HK systems with Euclidean confidence balls:
\[
 \sum_{i=1}^n\sum\nolimits_{t=0}^{\infty}\|x_i(t+1)-x_i(t)\|_2^2\leq \frac{n^2\rho_*^2}{4}<\infty,
\]

Our construction shows that no analogous estimate can hold for arbitrary non-Euclidean confidence norms.
Indeed, in Example~\ref{ex:two}, the moving agent alternates between $\ee$ and $2\ee$, so
\begin{equation}\label{eq:K-inf}
\sum\nolimits_{t=0}^{\infty}
\|x_5(t+1)-x_5(t)\|_2^2
=\sum\nolimits_{t=0}^{\infty}3=\infty.
\end{equation}
The divergence persists if the Euclidean norm in~\eqref{eq:K-inf} is replaced by any other norm,
not necessarily the one defining the confidence neighborhoods. More generally, along any nonconstant periodic trajectory, the kinetic $s$-energy introduced in~\cite{Chazelle2011} is infinite for every $s>0$. The same holds for the corresponding total $s$-energy computed over the undirected symmetrization of the confidence graph: the moving agent repeatedly trusts the stationary agent at the origin while remaining at a positive distance from it.

As our results demonstrate, the choice of confidence norm can fundamentally affect
the asymptotic behavior of heterogeneous HK dynamics. In particular, departing from Euclidean
geometry can change convergence properties of the model.

\subsection{Open problems}

\paragraph{Admissible confidence norms.} An interesting open problem is to characterize the norms for which the escape criterion in Proposition~\ref{prop:escape} can be satisfied.
Our results establish this property for non-Euclidean $\ell_q$ norms in dimensions $d\ge3$ and their invertible linear transformations. However, the class of norms satisfying the escape criterion is not restricted to invertible linear transformations of $\ell_q$ norms. The following simple extensions of Theorem~\ref{thm:ellq} illustrate this fact.
\begin{lemma}[Maxima of confidence norms]
\label{lem:max-norms}
Let $d\ge3$, let $T\in\R^{d\times d}$ be
invertible, and consider the norm
\[
\|x\|=\max_{1\le j\le k}\alpha_j\|Tx\|_{q_j},
\qquad \alpha_j>0.
\]
Suppose that either $1\le q_j<2$  for all $j$ or $2<q_j\le\infty$ for all $j$.
Then the escape criterion in
Proposition~\ref{prop:escape} can be
satisfied for every $m\ge2$.
\end{lemma}
\begin{proof}
By an invertible change of coordinates,
we may assume that $T=I$.
Choose $\ee$ and $w_i$ as in the proof of
Theorem~\ref{thm:ellq}, using the same sign
$\sigma$ for all $q_j$.

Since the family of exponents is finite,
for sufficiently large $K$ we have
\[
\|w_i-\ee\|_{q_j}>\|w_i\|_{q_j}
\]
for every $i=1,2,3$ and $j=1,\ldots,k$. Let $j_i$ attain the maximum defining
$\|w_i\|$. Then
\[
\|w_i-\ee\|
\ge\alpha_{j_i}\|w_i-\ee\|_{q_{j_i}}>\alpha_{j_i}\|w_i\|_{q_{j_i}}
=\|w_i\|.
\]
By permutation symmetry all norms $\|w_i\|$ coincide, and so do $\|w_i-\ee\|$.
The balancing condition is unchanged,
so Proposition~\ref{prop:escape} applies.
\end{proof}

For instance, in $\R^3$, Lemma~\ref{lem:max-norms} ensures that the norm
\[
\|x\|=\max\left\{\|x\|_3,\frac65\|x\|_\infty\right\}
\]
satisfies the escape criterion for every prescribed period. Its unit ball has both
flat facets and curved boundary portions, being neither strictly convex nor polyhedral.
Hence, this norm cannot be obtained from any $\ell_q$ ball by an invertible linear transformation.
A general characterization of admissible confidence norms remains open. 
Another example is given by the following remark. 
\begin{remark}[Another family of admissible norms]
Consider, in $\R^3$, the family of polyhedral norms
\[
\|x\|_*=\|x\|_1+
\alpha|x_1+x_2+x_3|,
\qquad 0<\alpha<1/3.
\]
These norms are not linearly isometric to any $\ell_q$ norm\footnote{Indeed, their unit balls have $14$ facets,
unlike the three-dimensional $\ell_1$ and $\ell_\infty$ balls, which have $8$ and $6$ facets, respectively; for
$1<q<\infty$, the $\ell_q$ balls are strictly convex.}. Nevertheless, the escape
criterion holds for every prescribed period. Indeed, let $H=2^{m-1}$, $\xi=5H-4$,
$\ee=(1,1,1)$, and
\[
w_1=(-1,-1,\xi+2),\quad
w_2=(-1,\xi+2,-1),\quad
w_3=(\xi+2,-1,-1).
\]
Then $\sum_{i=1}^3w_i=\xi\ee$, while $\|w_i-\ee\|_*-\|w_i\|_*=1-3\alpha>0.$
Thus Proposition~\ref{prop:escape} applies.

Geometrically, the construction can also be interpreted in $\ell_1$ on $\R^4$: the lifted vectors
$w_i$ have cyclically permuted first three coordinates and a common fourth coordinate
$\alpha\xi$, while $\ee$ lifts to $(1,1,1,3\alpha)$; all these vectors belong to the subspace
$\{y=(y^1,y^2,y^3,y^4):y^4=\alpha(y^1+y^2+y^3)\}$.

More generally, the same construction applies to
norms of the form
\[
\|x\|_*=\|x\|_1+\sum\nolimits_{j=1}^s\alpha_j p_j(x),
\]
where $p_j$ are seminorms invariant under the cyclic shift $(x_1,x_2,x_3)\mapsto(x_3,x_1,x_2)$, $\alpha_j\ge0$,
and $\sum\nolimits_{j=1}^s\alpha_j p_j(\ee)<1$. Indeed, the triangle inequality yields
$
\|w_i-\ee\|_*-\|w_i\|_*
\ge 1-\sum_{j=1}^s\alpha_j p_j(\ee)>0.
$
By cyclic invariance, the values $\|w_i\|_*$ are identical for $i=1,2,3$, as are the values
$\|w_i-\ee\|_*$. Hence a common radius $r$ satisfying~\eqref{eq:escape-norm} exists,
and the escape criterion remains valid for every prescribed period.
\end{remark}

We conjecture that, in dimensions $d\ge3$, the escape criterion can be satisfied for every
norm that is not induced by an inner product (equivalently, whose unit ball is not an ellipsoid).
Establishing this conjecture, however, remains an open problem.

\paragraph{Minimal configurations admitting nonconstant periodic trajectories.}
Our results establish that five is the minimum number of agents required for nonconstant periodic trajectories in heterogeneous
HK dynamics, and that this bound is attained in dimensions $d\ge3$ for suitable non-Euclidean confidence norms.
It can be shown (see Lemma~\ref{lem:frozen-vertices} in the Appendix) that every vertex of the opinion convex hull must remain occupied by a stationary agent along a periodic trajectory. Thus, the number of potentially moving agents is strongly constrained by the affine dimension of this hull.
Several questions remain open:
\begin{enumerate}
    \item Can five agents exhibit a nonconstant periodic trajectory in the scalar heterogeneous HK model? The example in~\cite{EtesamiTouri2026} shows that six agents suffice, while we prove that $n\le4$ agents are insufficient. 

    \item Can five agents exhibit nonconstant periodic trajectories in a two-dimensional opinion space,
    under some confidence norm? Proposition~\ref{prop:escape-dimension} shows that our escape mechanism cannot work in a normed plane,
    but does not exclude other periodic mechanisms. Lemma~\ref{lem:frozen-vertices} implies that, along any periodic trajectory with a noncollinear opinion hull,  at least three agents must remain stationary. Consequently, at most two agents can move in a planar five-agent system (collinear trajectories reduce to the scalar case of item 1). Moreover, for the Euclidean norm, the convergence theorem of Chazelle and Wang~\cite{ChazelleWang2017} implies that any such nonconstant periodic trajectory must involve exactly two moving agents with different
    confidence radii.

    \item Can five-agent periodic trajectories arise through switching patterns different from the mechanism studied here? 
    In particular, can one construct periodic trajectories with four stationary agents and
    one moving agent following a different sequence of confidence neighborhoods, or with three stationary agents
    and two moving agents? The latter possibility is also relevant to Euclidean  confidence norms, for which our escape criterion
    cannot be satisfied.

    \item How many distinct confidence radii are necessary for nonconstant periodic trajectories in the scalar
    heterogeneous HK model? The six-agent example of Etesami and Touri~\cite{EtesamiTouri2026} uses three distinct positive radii, whereas our multidimensional construction requires only two. It remains open whether two distinct positive confidence radii suffice for periodic behavior in the scalar model, regardless of the number of agents. A related question concerns noncollinear periodic trajectories in two-dimensional HK models with only two distinct positive radii. The question is also unresolved for Euclidean confidence norms in higher dimensions. Indeed, the convergence theorem of Chazelle and Wang~\cite{ChazelleWang2017} covers the special case where all non-stubborn agents have a common radius and does not extend to two distinct strictly positive radii.
\end{enumerate}

\section{Conclusion}\label{sec:conclusion}

We have shown that heterogeneous Hegselmann--Krause dynamics with multidimensional opinions can exhibit
periodic trajectories of arbitrarily large minimal period, even with only five agents and two distinct
positive confidence radii. The construction applies to all non-Euclidean $\ell_q$ norms in dimensions
$d\ge3$, as well as to broader classes of confidence norms. Remarkably, only one agent moves, and its
trajectory is collinear; the multidimensional nature of the phenomenon is entirely due to the
geometric arrangement of the stationary agents and their confidence neighborhoods.

We have also established that five agents are necessary for periodic behavior, regardless of
the dimension and confidence norm. In fact, trajectories with at most three agents always
converge exponentially, whereas four agents cannot exhibit nonconstant periodic orbits.
The geometric obstruction to our construction in dimensions $d\le2$ highlights the importance
of the geometry of confidence balls. Whether other mechanisms can generate periodic behavior
with five agents in scalar or planar opinion spaces, and which norms admit our escape
mechanism, remain interesting open questions.

\section{Declaration of AI-Assisted Tools}

During the preparation of this manuscript, the authors used GPT-6 (OpenAI) and Claude 5.5 Opus (Anthropic) to polish the phrasing of the text and to assist in drafting initial strategies and proofs for the Theorems~\ref{thm:ellq},~\ref{thm:n3-exponential} and~\ref{thm:n4}. The final proofs presented in the manuscript were rewritten and verified by the authors, who assume full responsibility for the contents and correctness of this manuscript.

\bibliographystyle{plain}
\bibliography{n5_multidimensional_references}

\appendix
\section{Proof of Theorem~\ref{thm:n4}}\label{app:n4}

We use the following standard fact repeatedly.
\begin{lemma}[Fixed neighborhoods]\label{lem:fixed-neighborhoods}
If all confidence neighborhoods $N_i(t)$ are constant along a periodic trajectory, the trajectory is
stationary.
\end{lemma}
\begin{proof}
Stack the opinions $x_i^{\top}$ as the rows of a matrix $X(t)\in\R^{n\times d}$. Then $X(t+1)=PX(t)$, where $P$ is a fixed
row-stochastic matrix with positive diagonal (since $i\in N_i$). By the Gershgorin circle theorem, $1$ is the only eigenvalue of $P$ on the unit
circle, and it is semisimple because the powers $P^k$ are bounded. Hence $P^k$ converges as
$k\to\infty$~\cite{ProskurnikovTempo2017}. A periodic sequence $P^tX(0)$ that converges is constant.
\end{proof}

\paragraph{Step 1: vertices are frozen.}

\begin{lemma}[Frozen vertices]\label{lem:frozen-vertices}
Along a periodic trajectory, with any norm and any number of agents, the convex hull $C(t)={\rm conv}\,\{x_1(t),\ldots,x_n(t)\}$ of the opinions is constant $C(t)\equiv C$. For each vertex $v$ of $C$, the set $V_v(t)=\{i:x_i(t)=v\}\ne\emptyset$ is constant; its agents are stationary and trust only agents in $V_v$.
\end{lemma}
\begin{proof}
Every new opinion is a convex combination of current opinions, so $C(t+1)\subseteq C(t)$, and periodicity
gives equality. A vertex $v$ of $C$ is one of the opinions, so $V_v(t)\ne\emptyset$. If $i\in V_v(t+1)$,
then $v$ is the average of the opinions $x_j(t)\in C$, $j\in N_i(t)$. Since $v$ is an extreme point of
$C$, all of them equal $v$, in particular, $x_i(t)=v$, that is, $i\in V_v(t)$. We have thus proved that 
$V_v(t+1)\subseteq V_v(t)$, and periodicity forces $V_v(t)$ to be constant. Furthermore, 
$N_i(t)\subseteq V_v$ (and hence $N_i(t)\equiv V_v$) for each $i\in V_v$. 
\end{proof}

\paragraph{Step 2: noncollinear hull.}
If $C$ is a single point, the trajectory is stationary. If $C$ is neither a point nor a segment, it has
at least three vertices, so by Lemma~\ref{lem:frozen-vertices} at least three agents are stationary. The
remaining agent, if any, moves among at most three fixed opinions. By
Proposition~\ref{prop:three-anchor-convergence} below, its opinion converges, and since it is periodic,
it is constant. This contradicts our assumption.

\begin{proposition}[One agent and three fixed opinions]\label{prop:three-anchor-convergence}
Let $x_1,x_2,x_3\in\R^d$ be fixed, not necessarily distinct, and the fourth agent with radius $\rho_4:=r>0$ 
updates its opinion $x_4(t):=z(t)$  in accordance with 
\[
z(t+1)=\frac{1}{|S(t)|}\sum_{q\in S(t)}x_q,\qquad S(t)=N_4(t)=\bigl\{q\in\{1,2,3,4\}:\|x_q-z(t)\|\le r\bigr\}.
\]
Then $S(t)$ is eventually constant and $z(t)$ converges as $t\to\infty$.
\end{proposition}
\begin{proof}
\emph{Observation 1. If $S(t)=\{1,2,3,4\}$ and $i\notin S(t+1)$, then $i\notin S(\tau)$ for all $\tau\geq t+1$.}

Without loss of generality, assume that $i=1$. Also, shifting the opinions, we may assume that $z(t)=0$, so that
$\|x_1\|,\|x_2\|,\|x_3\|\leq r$ while $w=z(t+1)=(x_1+x_2+x_3)/4$ satisfies the inequality $\|w-x_1\|>r$.

We will prove that $\|x_1-y\|>r$ for every $y\in K:=\operatorname{conv}\{w,x_2,x_3\}$. Indeed, write
\[
y=\lambda w+\mu x_2+\nu x_3=\frac{\lambda}{4}x_1+\left(\frac{\lambda}{4}+\mu\right)x_2+\left(\frac{\lambda}{4}+\nu\right)x_3,
\qquad
\lambda,\mu,\nu\ge0,\quad
\lambda+\mu+\nu=1.
\]
Using $4w=x_1+x_2+x_3$ and $\lambda+\mu+\nu=1$, it can be checked that
\[ 
\begin{aligned} x_1-y &=x_1-\lambda w-\mu x_2-\nu x_3\\ &=x_1-\lambda w-(1-\lambda)(x_2+x_3)   +\nu x_2+\mu x_3\\ &=x_1-\lambda w-(1-\lambda)(4w-x_1)   +\nu x_2+\mu x_3\\ &=(4-3\lambda)(x_1-w)-2(1-\lambda)x_1   +\nu x_2+\mu x_3, 
\end{aligned} 
\]
entailing that $\|x_1-y\|>(4-3\lambda)r-2(1-\lambda)r-\nu r-\mu r=r$. Since $w=z(t+1)\in K$ and $K$ is invariant
under averaging with any subset of $\{x_2,x_3\}$, the moving opinion remains
in $K$ for all subsequent times. Thus, agent $1$ is never trusted by agent $4$ again.

\emph{Observation 2. If $|S(t)|\le3$, then $S(t)\subseteq S(t+1)$.} Indeed, suppose that $S(t)=\{4,i,j\}$. Then
\[
z(t+1)=\frac{z(t)+x_i+x_j}{3}\Longrightarrow 
\|x_i-z(t+1)\|
\le\frac{2\|x_i-z(t)\|+\|x_j-z(t)\|}{3}
\le r,
\]
and similarly for $j$. The cases $|S(t)|<3$ are straightforward.

\emph{Conclusion.}
By Observation 2, the neighborhood can lose an agent only when $S(t)=\{1,2,3,4\}$.
By Observation 1, any such loss is irreversible, so the full neighborhood
can never reappear. Thereafter, the neighborhood is nondecreasing and hence
eventually constant. Thus, $S(t)=S$ for all sufficiently large $t$. 

If $S=\{4\}$, agent $4$ is stationary. Otherwise, it converges exponentially fast, since
\[
z(t+1)=\frac{z(t)+\sum_{q\in S\setminus\{4\}}x_q}{|S|}. \qedhere
\]
\end{proof}

\paragraph{Step 3: Reduction of the collinear case.}
Let the convex hull have affine dimension $1$, that is, $C=[p,q]$ with $p\ne q$. Writing $x_i(t)=p+\xi_i(t)(q-p)$, we have
$\|x_i(t)-x_j(t)\| =|\xi_i(t)-\xi_j(t)|\,\|q-p\|$. Thus, the scalars $\xi_i(t)\in[0,1]$ follow the scalar HK dynamics with confidence radii
$\rho_i/\|q-p\|$. Without loss of generality, we may therefore assume $d=1$ and $C=[0,1]$.

By Lemma~\ref{lem:frozen-vertices}, the agents at $0$ and $1$ are stationary and trust only
agents at the same point. Moreover, by Proposition~\ref{prop:three-anchor-convergence},
a periodic trajectory with at most one moving agent must be stationary.

Hence, any nonconstant periodic trajectory has exactly two stationary agents (\emph{anchors}),
one at $0$ and one at $1$, and two moving agents (\emph{movers}) with opinions $a(t),b(t)\in(0,1)$
and confidence radii $r_a,r_b>0$. 

In what follows, we use the distance between the movers $d(t):=|a(t)-b(t)|$.

\paragraph{Step 4: the movers stay on one side of $1/2$.}
The squares
\[
Q_-=[0,\tfrac12]^2,\qquad
Q_+=[\tfrac12,1]^2
\]
are forward invariant. Indeed, if $a,b\le1/2$, an agent trusting anchor $1$ necessarily trusts everybody and updates to
$(1+a+b)/4\le1/2$; all other updates average values not exceeding $1/2$.
The case of $Q_+$ follows by reflection. By periodicity, a trajectory meeting either
square must remain there at all times.

Suppose that the trajectory never enters $Q_-\cup Q_+$. The movers must then lie on
opposite sides of $1/2$. They cannot exchange sides in one step as this would require\footnote{Indeed, a mover with opinion $y<\frac12$ whose partner keeps opinion $z>\frac12$ can move above $\frac12$ only by updating its opinion to $(y+z)/2$ or $(1+y+z)/4$, both of which require $y+z>1$; symmetrically, the upper mover needs $y+z<1$.}
simultaneously $a+b>1$ and $a+b<1$. Hence, after relabeling, we may assume that
$a(t)<\tfrac12<b(t)$ throughout the trajectory.

We claim that agent $a$ always trusts anchor $0$ but never anchor $1$.
Indeed, if it does not trust $0$, then $r_a<a<1/2$, so it cannot trust
$1$ either. Its opinion is then nondecreasing (averaging with $b(t)$ and/or itself) and the link to $0$
cannot reappear. Periodicity makes $a$ stationary, contradicting
Proposition~\ref{prop:three-anchor-convergence}.
On the other hand, if $a$ trusts $1$,
it trusts everybody and updates to $(1+a+b)/4>a$. The link to $1$ then
persists, making $a$ strictly increasing, again a contradiction.
Symmetrically, agent $b$ always trusts $1$ but never $0$. Consequently, depending on whether the
movers trust each other,
\[
a(t+1)\in\left\{\frac a2,\frac{a+b}{3}\right\},
\qquad
b(t+1)\in\left\{\frac{1+b}{2},
                 \frac{1+a+b}{3}\right\}.
\]
Since $a<1/2<b$, we have
\[
a(t+1)\le\frac{a+b}{3},
\qquad
b(t+1)\ge\frac{1+a+b}{3},
\qquad
d(t+1)=b(t+1)-a(t+1)\ge\frac13.
\]
By periodicity, $d(t)\ge1/3$ at every time.

If the movers trust each other mutually at time $t$, then $d(t+1)=1/3$. Since mutual trust requires $\min\{r_a,r_b\}\ge d\ge1/3$,
it persists forever, and all confidence neighborhoods are constant in view of periodicity. 
Lemma~\ref{lem:fixed-neighborhoods} then excludes a nonconstant periodic orbit.

Otherwise, the mover with the smaller radius never trusts the other mover.
It therefore averages only with itself and its nearest anchor, moving strictly
toward that anchor. This again contradicts periodicity.

Thus, after reflection if necessary, we may assume that the trajectory lies in $Q_-$.
Moreover, a coordinate strictly below $1/2$ remains strictly below $1/2$.
Hence, on a periodic trajectory, any mover attaining $1/2$ must remain stationary,
which is excluded by Proposition~\ref{prop:three-anchor-convergence}.
We may therefore assume
\[
0<a(t),b(t)<1/2\quad\forall t=0,1,2\ldots
\]

\paragraph{Step 5: links to the anchors are constant.}

We show that, along a periodic trajectory, each mover either always trusts a given anchor
or never trusts it. It suffices to prove this for the mover with opinion $a$, since
the argument for $b$ is identical. The case $r_a\geq 1$ is trivial: in this case, mover $a$ trusts both anchors.

\emph{Case $r_a<1/2$.} The mover never trusts anchor $1$. Moreover,
once it trusts anchor $0$, this link persists. Indeed, if $a\le r_a$, then its next opinion is
either $a/2$ or $(a+b)/3$. In the latter case, $b\le a+r_a$, so
\[
a(t+1)=\frac{a(t)+b(t)}{3}
\le\frac{2a(t)+r_a}{3}\le r_a.
\]

\emph{Case $1/2\leq r_a<1$.} The mover $a$ always trusts anchor $0$ and the
other mover $b$. Denote $\theta=1-r_a\in(0,1/2]$.
Anchor $1$ is trusted precisely when $a\ge\theta$,
and the update is
\[
a(t+1)=
\begin{cases}
(a(t)+b(t))/3,&a<\theta,\\
(1+a(t)+b(t))/4,&a\ge\theta.
\end{cases}
\]
For this link to appear, we must have
\[
a(t)<\theta\le\frac{a(t)+b(t)}{3}\Longrightarrow \theta<\frac{\theta+\frac12}{3},
\]
which implies $\theta<1/4$. However, the trust of $a$ in anchor $1$, once acquired,
persists, because
\[
a(t+1)=\frac{1+a(t)+b(t)}{4}>\frac14>\theta.
\]
Thus, if $\theta\ge1/4$, the link to anchor $1$ cannot appear, whereas if
$\theta<1/4$, the link cannot disappear once established. Hence every link to an
anchor can change at most once and, by periodicity, must be constant.

\paragraph{Step 6: the links between the movers are constant.}

\emph{Case $r_a,r_b<\frac12$.} Neither mover trusts anchor $1$, and anchor $1$ trusts only itself
by Lemma~\ref{lem:frozen-vertices}. Hence anchor $0$ and the two movers form an autonomous
three-agent HK system. Its trajectory is periodic, so by Theorem~\ref{thm:n3-exponential} it
converges and is therefore constant, a contradiction.

\emph{Case $r_a,r_b\ge\frac12$.} Since $d=|a-b|<\frac12$, the movers always trust each other.
Together with Step~5, all confidence neighborhoods are constant.

\emph{Case $r_a\ge1/2>r_b$ (up to a relabelling).} By Step~5, agent $a$ always trusts anchor $0$
and agent $b$, and agent $b$ never trusts anchor $1$. If $a$ never trusts anchor $1$ either, the
previous reduction to a three-agent system applies. Hence $a$ always trusts anchor $1$. Let
$\delta\in\{0,1\}$ indicate whether $b$ trusts anchor $0$ (constant by Step~5), and let
$\eta(t)\in\{0,1\}$ indicate whether $b$ trusts $a$. Then
\[
a(t+1)=\frac{1+a(t)+b(t)}{4},\qquad
b(t+1)=
\begin{cases}
\frac{a(t)+b(t)}{2+\delta},&\eta=1,\\[2mm]
\frac{b(t)}{1+\delta},&\eta=0.
\end{cases}
\]

Suppose that $\eta(t)=1$ and $\eta(t+1)=0$, and put $s=a(t)+b(t)\in(0,1)$. Then
$a(t+1)=(1+s)/4$ and $b(t+1)=s/(2+\delta)$, so $a(t+1)>b(t+1)$ and
\[
d(t+1)=
\begin{cases}
(1-s)/4,&\delta=0,\\
(3-s)/12,&\delta=1.
\end{cases}
\]
As long as $\eta=0$, we have $d(\tau+1)=\bigl(d(\tau)+\gamma\bigr)/4$ with $\gamma=1-(1-\delta)s$.
Indeed, for $\delta=0$ the opinion $b$ stays at $s/2$, so
$d(\tau+1)=\bigl(d(\tau)+1-2b\bigr)/4$. For $\delta=1$ it is halved at each step, so
$d(\tau+1)=\bigl(1+d(\tau)\bigr)/4$. Since $0<d(t+1)<\gamma/3$, the distance increases
monotonically towards $\gamma/3$ while $\eta=0$. Moreover, $d(t+1)>r_b$, because the link is lost
at time $t+1$. Hence the link never reappears. On a periodic trajectory this is impossible unless
$\eta$ is constant, so again all confidence neighborhoods are constant. 

\paragraph{Step 7: conclusion.}
Every case not already excluded in Steps 2--6 yields constant confidence neighborhoods: persistent mutual trust in Step~4, and the second and third cases of Step~6. Lemma~\ref{lem:fixed-neighborhoods} then
shows that the trajectory is stationary, a contradiction. This proves Theorem~\ref{thm:n4}.\hfill\qed

\begin{remark}[Open confidence balls]\label{rem:n4-open}
Theorem~\ref{thm:n4} also holds for open confidence
neighborhoods
\[
N_i(t)=\{j:\|x_j(t)-x_i(t)\|<\rho_i\}.
\]
The proof requires only minor adjustments at the confidence thresholds.
In Proposition~\ref{prop:three-anchor-convergence},
if an initially trusted agent $1$ becomes untrusted,
then $\|x_1-w\|\ge r$, whereas $\|x_1\|,\|x_2\|,\|x_3\|<r$.
The same triangle-inequality argument gives
$\|x_1-y\|\ge r$ for every $y\in K$, ensuring
that the lost link cannot reappear.
Observation~2 remains valid with strict inequalities.

In Step~4, mutual trust requires $\min\{r_a,r_b\}>d(t)\ge1/3$.
Since mutual trust yields $d(t+1)=1/3$, it persists thereafter.
In Step~5, a link to anchor $0$ persists with strict inequalities. For the link to anchor
$1$, trust now requires $a>\theta$. A switch from distrust to trust therefore requires
\[
a\le\theta<\frac{a+b}{3}
<\frac{\theta+1/2}{3},
\]
which still implies $\theta<1/4$. Once acquired, this link remains present.

Finally, in Step~6, loss of the link between
the movers means $|a-b|\ge r_b$.
Since their separation subsequently increases,
the link cannot reappear.
The remaining arguments are independent of
whether the confidence balls are open or closed.
\end{remark}

\end{document}